\documentclass[aps,prc,twocolumn]{revtex4-1}

\usepackage{graphicx}
\usepackage{dcolumn}
\usepackage{bm}
\usepackage{amsmath}
\usepackage{amssymb}
\usepackage{isotope}
\usepackage{color}
\usepackage[colorlinks=true,linkcolor=blue,citecolor=blue,urlcolor=blue]{hyperref}

\begin{document}

\title{On the multipole mixing ratio of the 365 keV $ \boldsymbol{\gamma} $-transition in \textsuperscript{129}Ba}

\author{S. Chakraborty}
\email{mscphys@gmail.com}
\affiliation{Experimental Nuclear Physics Division, Physics Group, Variable Energy Cyclotron Centre, Kolkata, India.}

\begin{abstract}
The $ E2/M1 $ multipole mixing ratio ($ \delta $), reported by J. Gizon \textit{et al.} [\href{https://doi.org/10.1103/PhysRevC.17.596}{Phys. Rev. C 17, 596 (1978)}], for 365 keV $ \gamma $-transition in \isotope[129]Ba is reevaluated and found altered. Experimentally determined angular distribution coefficients indicate a large \textit{E}2 admixture in the 365 keV $ \gamma $-ray. Indeed, it is found in better agreement with the reported linear polarization value for this $ \gamma $-ray.
\end{abstract}

\maketitle

In Ref.~\cite{129Ba}, J. Gizon and co-workers measured the angular distribution and linear polarization of some $ \gamma $-rays in \isotope[129]Ba. In particular, for 365 keV $ \gamma $-transition, the following spectroscopic results were reported:
\begin{center}
A\textsubscript{22} = -- 0.70 (3); A\textsubscript{44} = 0.16 (5); P = 0.072 (68)
\end{center}
Based on the angular distribution (linear polarization) results, the multipole mixing ratio $ -0.97 \leqslant \delta \leqslant -0.79 $ ($ -0.84 \leqslant \delta \leqslant -0.31 $) was determined for this $ \gamma $-ray and finally, an effective value of $ -0.84 \leqslant \delta \leqslant -0.79 $ was adopted. However, the experimentally measured linear polarization also satisfies a higher magnitude of $ \delta $, as shown in the \figurename~2 of Ref.~\cite{129Ba}. In this context, the authors argued that `the region of overlap corresponding to an almost pure \textit{E}2 is rejected on the basis of angular distribution' \cite{129Ba}. Thus, the angular distribution results are re-examined to verify the aforementioned argument.

\begin{figure}[!h]
\centering
\includegraphics[trim=0cm 0cm 0cm 0cm,width=\columnwidth]{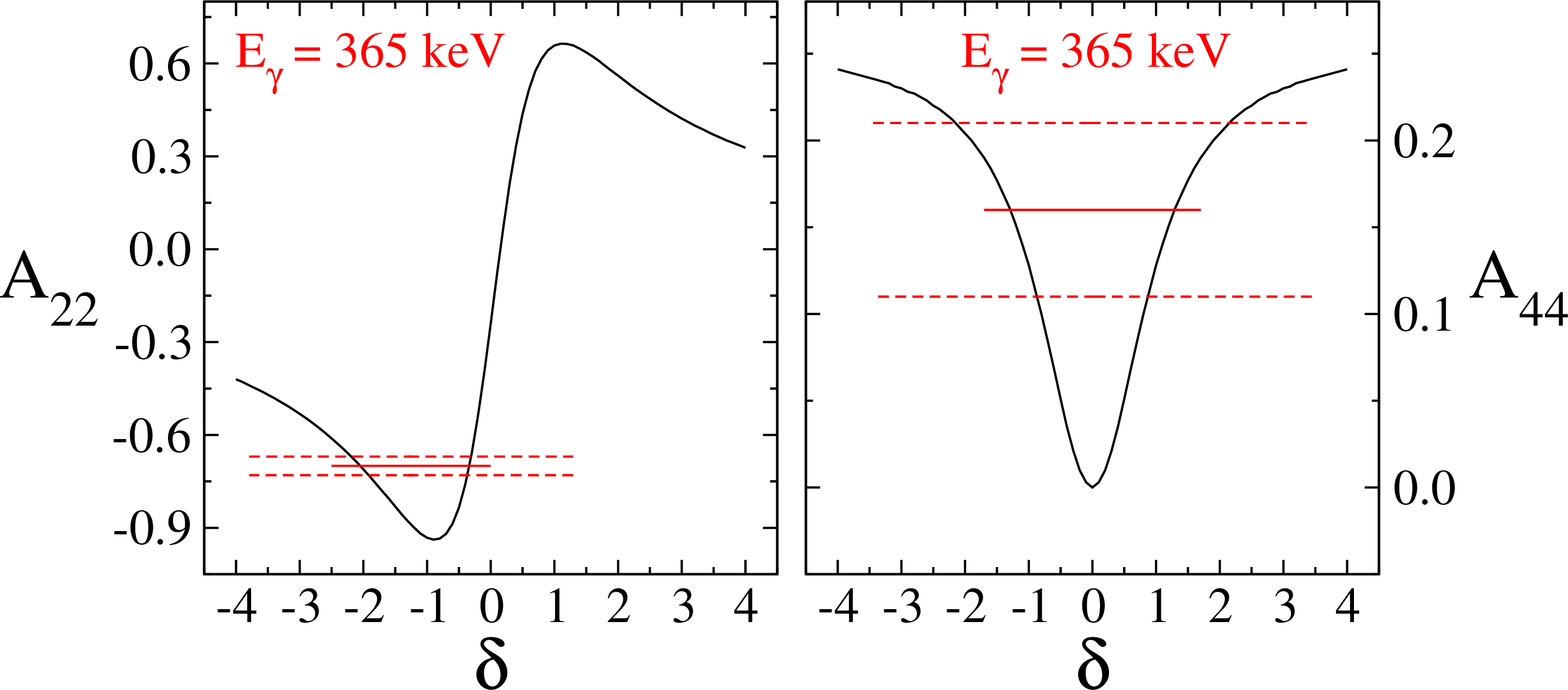}
\caption{Variation of theoretical angular distribution coefficients (black line), A\textsubscript{22} and A\textsubscript{44}, as a function of the multipole mixing ratio ($ \delta $). The experimental value (error) of A\textsubscript{22} and A\textsubscript{44} is shown in red solid (dashed) line.}
\label{AD}
\end{figure}

\begin{figure}[!h]
\centering
\includegraphics[trim=0cm 0cm 0cm 0cm,width=\columnwidth]{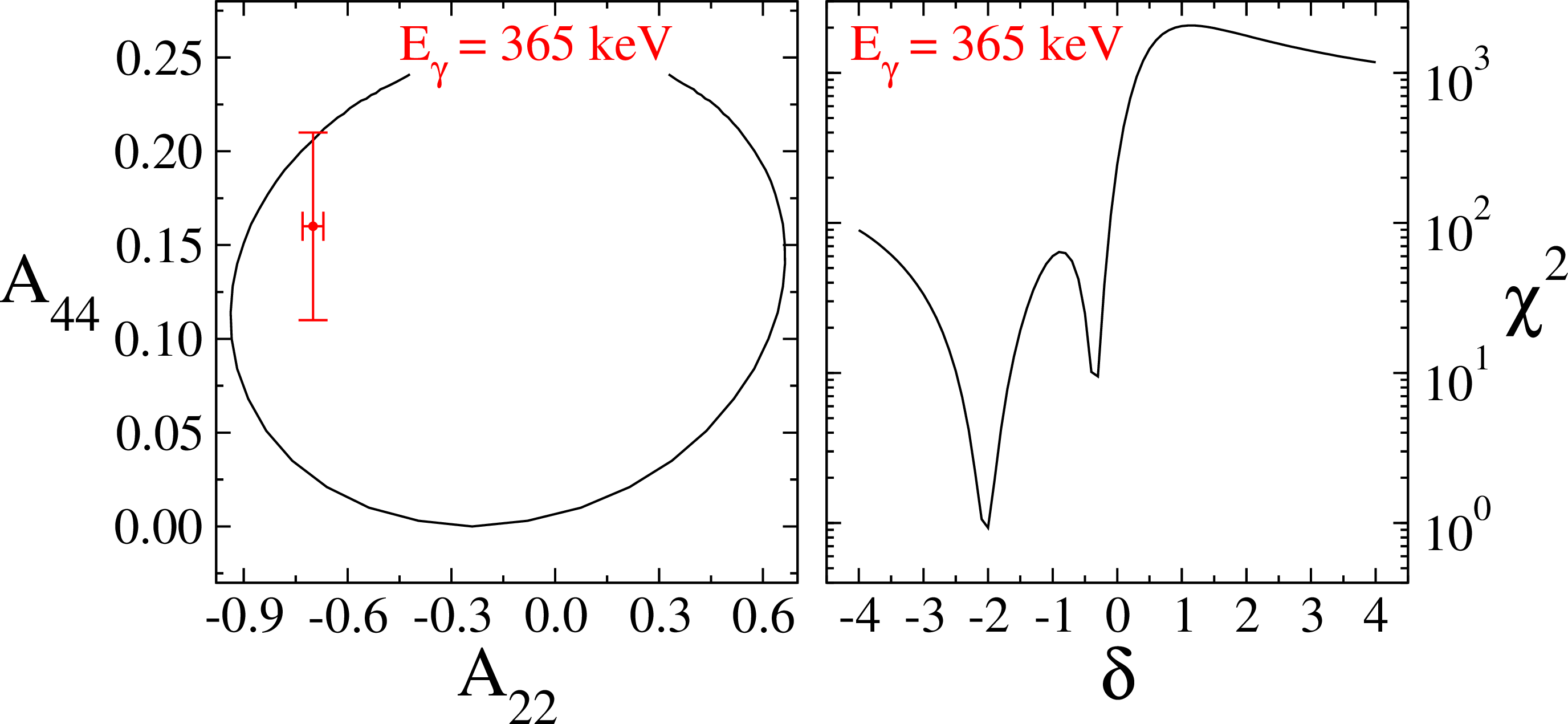}
\caption{Left: Contour plot of the angular distribution coefficients, A\textsubscript{22} versus A\textsubscript{44}, for different mixing ratios ($ \delta $). The experimental data point is marked in red color. Right: Plot of the $ \chi^{2} $ of A\textsubscript{22} and A\textsubscript{44} as a function of $ \delta $.}
\label{chi}
\end{figure}

\figurename~\ref{AD} shows that the reported values of angular distribution coefficient A\textsubscript{22} (A\textsubscript{44}) is possible for two different values of multipole mixing ratio $ \delta \approx $ -- 2.0 (2), -- 0.33 (3) ($ \delta \approx \pm 1.3~(7) $). However, the contour plot of A\textsubscript{22} versus A\textsubscript{44}, as presented in \figurename~\ref{chi}, clearly shows that these values are simultaneously possible only for the higher magnitude of $ \delta $, within the limit of uncertainty. Accordingly, the $ \chi^{2} $ of A\textsubscript{22} and A\textsubscript{44} is found to be minimum at $ \delta \approx -2.0 $  (\figurename~\ref{chi}). In the commented paper, $ \delta \approx -0.88 (9) $ ($ \delta \approx -0.58 (26) $), which corresponds to around 44(5)\% (25(16)\%) \textit{E}2 fraction, was estimated from angular distribution (linear polarization) measurement \cite{129Ba}. However, in the present analysis $ \delta \approx - 2.0 (2) $, which is equivalent to 80(4)\% \textit{E}2 contribution, is found for the 365 keV $ \gamma $-ray.

In the context of the present comment it is also worth noting that similar $ 13/2^{-} \rightarrow 11/2^{-} $ $ \gamma $-transition (E$ _{\gamma} $ = 483 keV) in the \textit{N} = 73 Xe isotone is also found highly mixed ($ \delta \approx -2.1 $) with about 81\% \textit{E}2 fraction \cite{127Xe}.

To summarize, earlier reported angular distribution result on 365 keV ($ 13/2^{-} \rightarrow 11/2^{-} $) $ \gamma $-ray in \isotope[129]Ba is revisited. Based on the present analysis, the 365 keV $ \gamma $-ray is found to be a predominant $ \Delta I = 1 $, \textit{E}2 transition.

SC is thankful to the Variable Energy Cyclotron Centre, Kolkata, India for Research Associateship \textit{vide} Ref. no. VECC/Admin/RA/Per-SC/2021/252.

\end{document}